\documentclass[reprint,
 amsmath,amssymb,
aps,
pra,
]{revtex4-2}
\usepackage{mathrsfs}
\usepackage{caption}
\usepackage{graphicx}
\usepackage{dcolumn}
\usepackage{bm}
\usepackage{float}
\usepackage{color}

\usepackage{hyperref}

\begin{document}

\preprint{APS/123-QED}

\title{$\mathcal{PT}$-Symmetric Spin-Boson Model with a Continuous Bosonic Spectrum: Exceptional Points and Dynamics} 

\author{Yong-Xin Zhang$^{1}$}

\author{Qing-Hu Chen$^{1,2,}$}
\email{qhchen@zju.edu.cn}
\affiliation{$^{1}$Zhejiang Key Laboratory of Micro-Nano Quantum Chips and Quantum Control, School of Physics, Zhejiang University, Hangzhou 310027, China\\
                $^{2}$Collaborative Innovation Center of Advanced Microstructures, Nanjing University, Nanjing 210093, China}
\date{\today}

\begin{abstract}
This work studies a $\mathcal{PT}$-symmetric non-Hermitian spin–boson model, consisting of a non-Hermitian two-level system coupled to a continuous bosonic bath. The static properties of the system are analyzed through a projection method derived from the displacement operator. We find that only a single exceptional point (EP) emerges, in contrast to non-Hermitian spin–boson models with finite modes, which typically exhibit multiple EPs. Notably, only a single real eigenvalue is found before the EP, which differs markedly from typical non-Hermitian systems where a pair of real eigenvalues precedes the EP. The time evolution of observables is further investigated via the Dirac–Frenkel time-dependent variational principle. Compared to its Hermitian counterpart, the non-Hermitian model exhibits distinct dynamical signatures, most notably the emergence of oscillations with periodic amplified amplitude. In the $\mathcal{PT}$-unbroken phase, the system exhibits sustained oscillatory dynamics with suppressed decoherence, whereas in the $\mathcal{PT}$-broken phase, additional dissipative channels accelerate decoherence and drive rapid convergence toward a stable steady state. These results shed light on how $\mathcal{PT}$ symmetry protects coherent light–matter interactions in non-Hermitian quantum systems.
\end{abstract}

\maketitle

\section{Introduction}

In realistic quantum systems, inevitable coupling to the environment necessitates a rigorous dynamical description within the framework of open quantum systems~\cite{Scully1997__,Weiss2012__,Meystre2021__}. Conventional theoretical tools, such as the quantum master equation~\cite{Gorini1976JMP_17_821-825,Lindblad1976CP_48_119-130}, can successfully describe dissipation and noise. However, they show clear limitations in high-dimensional or strongly coupled regimes. In recent years, effective theories based on non-Hermitian Hamiltonians have attracted increasing attention~\cite{Moiseyev_2011,Ashida2020AP_69_249-435}. These approaches incorporate dissipation and gain directly into non-Hermitian terms, providing a more concise and physically transparent framework to efficiently reveal key emergent phenomena in open systems.

Non-Hermitian Hamiltonians provide a new paradigm for studying open quantum systems~\cite{Mostafazadeh2002JMP_43_205-214,Mostafazadeh2002JMP_43_2814-2816a,Brody2013JPAMT_47_035305,Fring2017PRA_95_010102,Tzeng2021PRR_3_013015}. The non-Hermitian framework not only simplifies calculations, but also offers a novel physical perspective, challenging traditional notions rooted in closed Hermitian systems. Under this paradigm, numerous unprecedented phenomena have been predicted and observed, including exceptional point (EP)~\cite{El-Ganainy2018NP_14_11-19,Miri2019S_363_eaar7709,Ozdemir2019NM_18_783-798,Li2023NN_18_706-720,Zhang2025PRL_135_230203}, non-Hermitian decoherence~\cite{Castagnino2012JPAMT_45_444009,Gao2017PRE_95_013308,Duarte2018E_121_50006,Dey2019PLA_383_125931,Naghiloo2019NP_15_1232-1236,Bhat2023UJoP_68_101,Duttatreya2026AoP_484_170298a}, as well as other effects~\cite{Lee2016PRL_116_133903,Leykam2017PRL_118_040401,MartinezAlvarez2018PRB_97_121401a,Ochkan2024NP_20_395-401a}. These phenomena have been experimentally observed on various platforms, including optical waveguides~\cite{Guo2009PRL_103_093902,Ruter2010NP_6_192-195,Kottos2010NP_6_166-167,Zhang2019LSA_8_88}, cold atomic systems~\cite{Longhi2013PRA_88_062112,Lee2014PRX_4_041001,Li2019NC_10_855,Li2022PRL_129_093001}, and superconducting vortex systems~\cite{Hatano1996PRL_77_570-573,Feinberg1999PRE_59_6433-6443,dograQuantumSimulationParity2021}, demonstrating the broad applicability and potential of non-Hermitian physics. 

Notably, Bender \textit{et al.}~\cite{Bender1998PRL_80_5243-5246,Bender2002JPAMG_35_L467} demonstrated that a non-Hermitian Hamiltonian with parity-time ($\mathcal{PT}$) symmetry can have a purely real eigenvalue spectrum under specific conditions. This finding sparked extensive research into $\mathcal{PT}$-symmetric non-Hermitian extensions of the Rabi model~\cite{Rabi1936PR_49_324-328,Braak2016JPAMT_49_300301}, a key framework for light-matter interaction. Joglekar and Lee~\cite{Joglekar2014PRA_90_040101,Lee2015PRA_92_042103} considered a purely imaginary coupling constant within the semiclassical Rabi model~\cite{Xie2018PRA_98_052122,Liu2025PRR_7_023004}, revealing gain-loss dynamics in two-level systems and prompting further studies on imaginary qubit–cavity couplings in quantum Rabi models~\cite{Tian2023JPBAMOP_56_095001a,Ying2024AQT_7_2400288,Li2025AQT_8_2400609}. Recently, the introduction of a purely imaginary energy bias in quantum Rabi models has attracted growing interest, revealing the emergence and disappearance of EPs with varying light–matter coupling and unconventional dynamics~\cite{Lu2023PRA_108_053712,Wang2025PRA_112_043704}. Extended models~\cite{Naghiloo2019NP_15_1232-1236} further exhibit distinct decoherence behaviors in the $\mathcal{PT}$-unbroken and $\mathcal{PT}$-broken phases. Most recently, a dissipative qubit was introduced into a finite multi-mode spin-boson model, revealing distinctive dynamical behaviors characteristic of non-Hermitian systems~\cite{Zhang2024JCP_161_}.

To the best of our knowledge, the non-Hermitian spin–boson model with a continuous bosonic environment has rarely been studied. Introducing a purely imaginary bias into the standard spin–boson model gives it $\mathcal{PT}$ symmetry, effectively describing a $\mathcal{PT}$-symmetric qubit with balanced gain and loss~\cite{Naghiloo2019NP_15_1232-1236,Lu2023PRA_108_053712,Han2023PRL_131_260201}. Coupled to a continuous bosonic environment, the system features two dissipation channels, allowing us to study how this gain–loss balance influences coherent dynamics and relaxation. A natural question then arises: does such a model exhibit exceptional points, given that, aside from the lowest-energy eigenvalue, other eigenvalues may form a continuous band rather than discrete levels, similar to the Hermitian counterpart? How does non-Hermiticity modify the dynamical behavior of the system? The challenge further lies in the fact that even for the Hermitian spin–boson model, both analytical and numerical studies have long faced significant difficulties. Pioneering work in this area began with the Silbey–Harris (SH) variational method~\cite{Silbey1984JCP_80_2615-2617}, based on unitary polaron transformation. Subsequently, Alex \textit{et al.}~\cite{Chin2011PRL_107_160601} extended the original symmetric single coherent state to an asymmetric one, overcoming the limitations of the SH method. The SH scheme has since been extensively validated and further refined to higher orders ~\cite{leeAccuracySecondOrder2012a,heImprovedSilbeyHarrisPolaron2018,leboiteTheoreticalMethodsUltrastrong2020}. For dynamical studies, Wu \textit{et al.}~\cite{Wu2013JCP_138_084111,Duan2013JCP_139_044115} employed the Dirac–Frenkel time‑dependent variational principle (TDVP) with a similar trial state, demonstrating its effectiveness in capturing complex dynamical behaviors. However, the possible complex eigenvalues currently preclude a variational principle to determine the lowest eigenvalues in non-Hermitian systems~\cite{Zhong2025PRL_135_106502}. Moreover, the ground state is also hard to define.

Based on these advances, a projection method is employed for studying the non-Hermitian spin-boson model. By assigning asymmetric bosonic states to the upper and lower levels and projecting the eigenstate onto different orthogonal bases, we obtain a set of self-consistent equations that determine the eigensolutions of the model. Combined with the TDVP approach, the method can be further applied to investigate dynamical properties. Under the protection of $\mathcal{PT}$ symmetry, our results show that non-Hermitian systems exhibit clearly distinct behaviors in the $\mathcal{PT}$-unbroken and  $\mathcal{PT}$-broken phases, including variations in the static eigenvalue spectrum and in the dynamical evolution.  This framework is expected to bridge the gap in understanding both finite-mode and continuous spectrum as well as the dynamics of non-Hermitian spin-boson systems, thereby advancing our understanding of $\mathcal{PT}$-symmetry-protected non-Hermitian effects.

The paper is organized as follows. In Sec.~\ref{Model and Method}, we introduce the $\mathcal{PT}$-symmetric non-Hermitian spin–boson model and describe the theoretical methods employed. Sec.~\ref{Result and Discussion} presents and discusses our key findings on how $\mathcal{PT}$-symmetry influences the spectrum and dynamical properties of the non-Hermitian spin-boson model, as well as the differences from its Hermitian counterpart. Sec.~\ref{Conclusion} summarizes the main conclusions and offers an outlook for future research. In Appendix~\ref{Finite Number}, we present the properties of the non-Hermitian spin-boson model with finite modes. Appendix~\ref{Appen-H} focuses on the Hermitian spin–boson model.

\section{Model and Methods}
\label{Model and Method}
\subsection{Non-Hermitian Spin–Boson Model}
The non-Hermitian spin-boson model consists of a non-Hermitian qubit coupled to a continuous bosonic bath. Generally, the system can be represented by the Hamiltonian
\begin{equation}
H=-\frac{\Delta }{2}\sigma _{x}+i\frac{\epsilon }{2}\sigma
_{z}+\sum_{j}\omega _{j}a_{j}^{\dag }a_{j}+\sigma
_{z}\sum_{j}g_{j}(a_{j}+a_{j}^{\dag }), 
\label{H}
\end{equation}
where $\sigma _{i=x,z}$ are the Pauli matrices, $\Delta$ is the tunneling amplitude, and $\epsilon $ is the energy bias. The operators $a_{j}$ and $a_{j}^{\dag }$ represent the bosonic annihilation and creation operators associated with harmonic oscillators of frequency $\omega_{j}$. Each oscillator couples to the qubit with strength $g_{j}$. The properties of the harmonic bath are completely determined by the spectral density
\begin{equation}
J(\omega)=\sum_j g_j^2\delta(\omega_j-\omega)
= \tfrac{1}{2}\lambda\omega_c^{1-s}\omega^s,\; 0<\omega<\omega_c.
\label{J}
\end{equation}
Throughout the paper, we use an Ohmic spectral density $s=1$ with the cut-off frequency $\omega _{c}=1$. The dimensionless parameter $\lambda $ denotes the coupling strength between the qubit and the harmonic bath.

When $\epsilon =0$, the system possesses  $\mathbb{Z}_{2}$ symmetry, with the
corresponding parity-conserving operator given by $\mathcal{P}=\sigma
_{x}\,\otimes \,\exp [i\pi \sum\limits_{j}a_{j}^{\dag }a_{j}]$, whose eigenvalues are
 $\pm 1$. However, for $\epsilon \neq 0$, the $\mathbb{Z}_{2}$ symmetry is no longer preserved. Considering the standard time-reversal
operator $\mathcal{T}$, which acts as complex conjugation,
and noting that $\mathcal{P}a_{j}(a_{j}^{\dag })\mathcal{P}=-a_{j}(a_{j}^{\dag
})$, we find that the model in Eq.~(\ref{H}) satisfies the commutation relation $[\mathcal{PT},H]=0$. This indicates that the Hamiltonian is $\mathcal{PT}$-symmetric.

For the eigenstate $\left\vert \Psi \right\rangle $ of Hamiltonian with eigenvalue $E$, satisfying $H\left\vert \Psi \right\rangle =E\left\vert \Psi
\right\rangle $, the $\mathcal{PT}$-symmetric Hamiltonian leads to the relation $\mathcal{PT}H\left\vert \Psi \right\rangle =H\mathcal{PT}\left\vert \Psi \right\rangle =E^{\ast }\mathcal{PT}\left\vert \Psi\right\rangle $. This implies that $\mathcal{PT}\left\vert \Psi \right\rangle 
$ is also the eigenstate of $H$, with the eigenvalue $E^{\ast}$. Under specific conditions, when the system's energy spectrum is purely real (i.e., $E = E^{\ast}$), it follows that $\mathcal{PT}\left\vert \Psi\right\rangle =\left\vert \Psi \right\rangle $, which indicates that the system's eigenstate $\left\vert \Psi \right\rangle $ resides in the $\mathcal{PT}$-unbroken phase. If the energy spectrum is complex, then $\mathcal{PT}\left\vert \Psi \right\rangle \neq \left\vert \Psi \right\rangle $, and the system's eigenstate $\left\vert \Psi \right\rangle $ is in the $\mathcal{PT}$-broken phase. The boundary between these two phases is marked by an EP. In these EPs, not only do the eigenvalues degenerate but also the corresponding eigenvectors coalesce into a single vector. 

The non-Hermitian qubit in Eq.~\eqref{H} can be traced back to a passive $\mathcal{PT}$-symmetric qubit, which is described by
$H_\text{passive} = -\frac{\Delta}{2} \sigma_x + i \epsilon \sigma_+ \sigma_-$, which contains only unidirectional loss in the upper level. The standard $\mathcal{PT}$-symmetric qubit in Eq.~\eqref{H} is obtained by adding a constant imaginary shift $-i \frac{\epsilon}{2} I$ to the spectrum, effectively balancing gain and loss. In this sense, the standard $\mathcal{PT}$ qubit can be viewed as a shifted and balanced extension of the passive qubit, preserving the $\mathcal{PT}$-symmetric structure while enabling richer non-Hermitian phenomena. In particular, their spectrum and dynamics coincide when considering energy offsets and suitable wave-function renormalization~\cite{Lu2023PRA_108_053712}. 

In the following, we will explore related phenomena in the non-Hermitian spin--boson model to reveal its intrinsic non-Hermitian properties protected by the $\mathcal{PT}$-symmetry. Before proceeding, we first clarify that the system described in Eq.~\eqref{H}, modeled with continuous bosonic modes~\cite{bullaNumericalRenormalizationGroup2003,zhangQuantumPhaseTransition2010}, can be more clearly expressed as
\begin{equation}
\begin{aligned}
H &= -\frac{\Delta}{2}\sigma_x + i\,\frac{\epsilon}{2}\sigma_z 
+ \int_0^{1}\! d\omega\, f(\omega)\, a_\omega^\dagger a_\omega \\
&\quad + \sigma_z \int_0^{1}\! d\omega\, h(\omega)\, (a_\omega + a_\omega^\dagger).
\end{aligned}
\end{equation}
In the one-dimensional representation, the bosonic bath dispersion is characterized by $f(\omega)$, while the coupling between the spin and the bath is described by $h(\omega)$. These two energy-dependent functions are related to the spectral function $J(\omega)$ via
\begin{equation}
J(x) = \frac{d\omega(x)}{dx} \, h^2[\omega(x)], \quad x \in [0, \omega_c],
\end{equation}
where $\omega(x)$ is the inverse function of $f(x)$, i.e., $f[\omega(x)] = x$. The above function does not uniquely determine $f(x)$ and $h(x)$, but an appropriate choice of $h(x)$ can simplify the calculation.

\subsection{Projection Method for Determining Eigenstates}

\label{Projection Method}
In spectral calculations, the continuous spectral function $J(\omega)$ is discretized into a finite set of bosonic modes, and the continuum limit is reached once the results converge numerically. The spectral function is discretized using the Wilson scheme with a discretization parameter $\Lambda$~\cite{wilsonRenormalizationGroupCritical1975a,krishna-murthyRenormalizationgroupApproachAnderson1980,bullaNumericalRenormalizationGroup2003,zhangQuantumPhaseTransition2010}. 
Starting from the highest-energy interval $[\Lambda ^{-1}\omega
_{c},\,\omega _{c}]$, the procedure iterates over logarithmically shrinking energy intervals $[\xi_{k+1},\,\xi _{k}]$, where $\xi_{k}=\Lambda ^{-k}\omega _{c}$ and $\Lambda > 1$. This discretization maps the bath onto a set of discrete bosonic modes, resulting in the discrete Hamiltonian:
\begin{equation}
H = -\frac{\Delta}{2}\sigma_x + i\,\frac{\epsilon}{2}\sigma_z
+ \sum_{k=0}^{\infty} \xi_k a_k^\dagger a_k
+ \sigma_z \sum_{k=0}^{\infty} \gamma_k (a_k + a_k^\dagger),
\label{HD}
\end{equation}
where the discretized frequencies $\xi_k$ and couplings $\gamma_k$ are given by
\begin{subequations}
\begin{align}
\xi_k &= \frac{s+1}{s+2}\,\frac{1-\Lambda^{-(s+2)}}{1-\Lambda^{-(s+1)}} \,\omega_c \,\Lambda^{-k}, \\ \gamma_k^2 &= \frac{\lambda \,\omega_c^2}{2}\,\frac{1-\Lambda^{-(s+1)}}{s+1}\,\Lambda^{-k(s+1)}.
\end{align}
\label{D}
\end{subequations}

Inspired by previous studies~\cite{Silbey1984JCP_80_2615-2617,Chin2011PRL_107_160601,leeAccuracySecondOrder2012a,heImprovedSilbeyHarrisPolaron2018,leboiteTheoreticalMethodsUltrastrong2020}, we employ a generalized SH ansatz to solve the Schrödinger equation for the non-Hermitian spin--boson model, which is sufficient since the full many-body bath correlations are not our focus. The method is based on an explicit generalized SH ansatz expressed in the $\sigma_z$ basis as
\begin{equation}
\left\vert \Psi \right\rangle =\left(
\begin{array}{c}
D(\alpha _{k})\left\vert 0\right\rangle  \\
rD(\beta _{k})\left\vert 0\right\rangle
\end{array}%
\right),  \label{state}
\end{equation}%
where $D[\alpha _{k}]=\exp [\sum_{k}(\alpha _{k}a_{k}^{\dag }-\alpha
_{k}^{\ast}a_{k})]$ is the displacement operator. Here, $\alpha
_{k},\beta _{k}$ and $r$ are all complex. Substituting this into the Schrödinger equation $H\left\vert \Psi \right\rangle = E\left\vert \Psi
\right\rangle $ yields
\begin{subequations}
\begin{align}
\begin{split}
\left[ i\frac{\epsilon}{2}+\sum_{k}\xi_{k}a_{k}^{\dag}a_{k}
+\sum_{k}\gamma_{k}(a_{k}+a_{k}^{\dag})\right] D(\alpha_{k})\left\vert 0\right\rangle \\
-\frac{\Delta}{2}rD(\beta_{k})\left\vert 0\right\rangle =ED(\alpha_{k})\left\vert 0\right\rangle,
\end{split} \label{S1} \\
\begin{split}
-\left[ i\frac{\epsilon}{2}-\sum_{k}\xi_{k}a_{k}^{\dag}a_{k}
+\sum_{k}\gamma_{k}(a_{k}+a_{k}^{\dag})\right] rD(\beta_{k})\left\vert 0\right\rangle \\
-\frac{\Delta}{2}D(\alpha_{k})\left\vert 0\right\rangle =rED(\beta_{k})\left\vert 0\right\rangle.
\end{split} \label{S2}
\end{align}
\end{subequations}
We project both sides of Eq.~\eqref{S1} onto the orthogonal states 
$\langle 0|D^\dag(\alpha_k)$ and $\langle 0| a_k D^\dag(\alpha_k)$. 
Similarly, Eq.~\eqref{S2} is projected onto 
$\langle 0|D^\dag(\beta_k)$ and $\langle 0| a_k D^\dag(\beta_k)$. 
This procedure yields the following four coupled equations
\begin{subequations}
\begin{gather}
E=-\frac{\Delta \eta r}{2}+i\frac{\epsilon }{2}+\sum_{k}\xi_{k}|\alpha
_{k}|^{2}+\sum_{k}\gamma_{k}(\alpha _{k}+\alpha _{k}^{\ast }), \\
\frac{\Delta \eta r}{2}(\alpha _{k}-\beta _{k})+\xi_{k}\alpha
_{k}+\gamma_{k}=0, \\
E=-\frac{\Delta \eta ^{\ast }}{2r}-i\frac{\epsilon }{2}+\sum_{k}\xi_{k}|\beta _{k}|^{2}-\sum_{k}\gamma_{k}(\beta _{k}+\beta _{k}^{\ast }), \\
\frac{\Delta \eta ^{\ast }}{2r}(\beta _{k}-\alpha _{k})+\xi_{k}\beta
_{k}-\gamma_{k}=0.
\end{gather}
\label{E}
\end{subequations}
Here, the overlap between the coherent states associated with the spin-up
and spin-down configurations is given by $\eta =\left\langle 0\right\vert
D^{\dag }(\alpha _{k})D(\beta _{k})\left\vert 0\right\rangle =\exp [-\frac{1%
}{2}\sum_{k}(\left\vert \beta _{k}\right\vert ^{2}+\left\vert \alpha
_{k}\right\vert ^{2})]\exp (\sum_{k}\alpha _{k}^{\ast }\beta _{k})$. By solving the four coupled equations above self-consistently, one obtains the eigensolutions of the non-Hermitian spin-boson model. It should be noted that the present framework is tailored to characterize the low-energy spectral branch, rather than to reconstruct the complete continuous band structure.
 Unlike its Hermitian
counterpart, the eigenvalues of this model can be complex, $E=E_{\mathrm{R}}+iE_{\mathrm{I}}$.
When $E_{\mathrm{I}}\neq 0$, the $\mathcal{PT}$-broken phase emerges.

For the spectral calculations presented here, a finite number of modes $M=80$ is employed in Sec.~\ref{Energy Spectra}, which ensures convergence with $\Lambda=1.2$, i.e., further increasing the number of modes does not modify the low-energy spectral structure or the position of the EP, indicating that the results have effectively converged toward the continuous limit.

\subsection{Dynamical Approach Based on the Dirac-Frenkel Time-Dependent Variational Principle}

Since the complete spectrum is inaccessible, the time evolution, involving contributions from multiple excited states, cannot be obtained from a full spectral decomposition and must instead be calculated directly.

To fully capture the dynamical contributions from the time-evolving states of the system, we employ the homogeneous discretization method~\cite{stockSemiclassicalSelfconsistentfieldApproach1995}. The discretized Hamiltonian takes the same form as Eq.~\eqref{HD}, where the frequencies of the $M$ harmonic modes are uniformly distributed as $\xi_k = k\Delta\xi$ with spacing $\Delta\xi = \xi_{\max}/M$ and $\xi_{\min}=\xi_1=\Delta\xi$. The corresponding couplings satisfy $\gamma_k^2 = J(\xi_k)\Delta\xi$. For real-time dynamics, a smooth exponential cutoff is adopted in the spectral density to ensure well-behaved bath correlation functions, while $\xi_{\max}$ serves as a numerical upper bound chosen sufficiently larger than the physical cutoff scale $\omega_c$. The spacing $\Delta\xi $ defines the Poincaré recurrence time $T_{\text{p}}=2\pi /\Delta\xi $, which must exceed all relevant timescales.

Similarly to the static approach, the dynamical ansatz is constructed from coherent states with asymmetric time-dependent parameters, which corresponds to a dynamical extension of Eq. \eqref{state}
\begin{equation}
\left\vert \Psi (t)\right\rangle =\left(
\begin{array}{c}
l(t)D[\alpha _{k}(t)]\left\vert 0\right\rangle  \\
r(t)D[\beta _{k}(t)]\left\vert 0\right\rangle
\end{array}%
\right) .  \label{STATE}
\end{equation}%
Here,  $D[\alpha _{k}(t)]=\exp [\sum_{k}(\alpha _{k}(t)a_{k}^{\dag }-\alpha
_{k}^{\ast }(t)a_{k})]$ is the time-dependent displacement operator. The
quantities  $\alpha _{k}(t)$, $\beta _{k}(t)$ denote the time-dependent
complex displacements for the $k$-th photon mode, while $l(t)$ and $r(t)$
represent the corresponding time-dependent complex amplitudes of the spin-up
and spin-down components, respectively.

To derive the equations of motion for the evolution parameters, we adopt the
Lagrangian formalism within the TDVP~\cite%
{Wu2013JCP_138_084111,Duan2013JCP_139_044115}. The Lagrangian associated
with the trial state $\left\vert \Psi (t)\right\rangle $ is defined as
\begin{equation}
L=\left\langle \Psi (t)\right\vert \frac{i}{2}\frac{\overleftrightarrow{%
\partial }}{\partial t}-H\left\vert \Psi (t)\right\rangle .
\end{equation}

Substituting the ansatz of Eq.~\eqref{STATE} into this expression yields
\begin{equation}
\begin{aligned} L &= \frac{i}{2}\bigl[\,l^{\ast}(t)\dot{l}(t) -
\dot{l}^{\ast}(t)l(t) + r^{\ast}(t)\dot{r}(t) - \dot{r}^{\ast}(t)r(t)\bigr]
\\ &\quad + \frac{i}{2}\sum_{k}\Bigl\{
|l(t)|^{2}\bigl[\alpha_{k}^{\ast}(t)\dot{\alpha}_{k}(t) -
\dot{\alpha}_{k}^{\ast}(t)\alpha_{k}(t)\bigr] \\ &\qquad\qquad +
|r(t)|^{2}\bigl[\beta_{k}^{\ast}(t)\dot{\beta}_{k}(t) -
\dot{\beta}_{k}^{\ast}(t)\beta_{k}(t)\bigr]\Bigr\} \\ &\quad - \langle
\Psi(t)|H|\Psi(t)\rangle, \end{aligned}
\end{equation}%
where the expectation value of the Hamiltonian reads
\begin{equation}
\begin{aligned} &\langle \Psi(t) | H | \Psi(t) \rangle \\ &\quad = |l(t)|^2
\sum_{k} \bigl\{ \xi_k |\alpha_k(t)|^2 + \gamma_k\bigl[\alpha_k^*(t) + \alpha_k(t)\bigr]\bigr\}
 \\ &\qquad + |r(t)|^2 \sum_{k} \bigl\{ \xi_k |\beta_k(t)|^2 - \gamma_k
\bigl[\beta_k(t) + \beta_k^*(t)\bigr] \bigr\} \\ &\qquad - \frac{\Delta}{2} \bigl[
l^*(t) r(t) \eta(t) + r^*(t) l(t) \eta^*(t) \bigr] \\ &\qquad + i
\frac{\epsilon}{2} \bigl[ |l(t)|^2 - |r(t)|^2 \bigr]. \end{aligned}
\end{equation}%
The Dirac--Frenkel TDVP yields the equations of motion by means of the
Euler--Lagrange equation
\begin{equation}
\frac{d}{dt}\left( \frac{\partial L}{\partial \dot{u}_{n}^{\ast }}\right) -%
\frac{\partial L}{\partial u_{n}^{\ast }}=0,
\end{equation}%
where $u_{n}^{\ast }$ denotes the complex conjugate of the variational
parameters $u_{n}$, namely $l(t)$, $r(t)$, $\alpha (t)$, and $\beta (t)$.
Hereafter, we suppress the explicit time argument ($t$) for brevity. The
resulting equations of motion can then be written explicitly as
\begin{subequations}
\begin{align}
i\dot{l}& +\frac{\Delta }{2}\eta r-l\Biggl\{\frac{i\epsilon }{2}-\frac{i}{2}%
\sum_{k}(\alpha _{k}^{\ast }\dot{\alpha}_{k}-\dot{\alpha}_{k}^{\ast }\alpha
_{k})  \notag \\
& \quad +\sum_{k}\bigl[\xi_{k}|\alpha _{k}|^{2}+\gamma_{k}(\alpha _{k}+\alpha
_{k}^{\ast })\bigr]\Biggr\}=0, \\[6pt]
i\dot{r}& +\frac{\Delta }{2}\eta ^{\ast }l+r\Biggl\{\frac{i\epsilon }{2}+%
\frac{i}{2}\sum_{k}(\beta _{k}^{\ast }\dot{\beta}_{k}-\dot{\beta}_{k}^{\ast
}\beta _{k})  \notag \\
& \quad -\sum_{k}\bigl[\xi_{k}|\beta _{k}|^{2}-\gamma_{k}(\beta _{k}+\beta
_{k}^{\ast })\bigr]\Biggr\}=0, \\[6pt]
il\dot{\alpha}_{k}& -l(\xi_{k}\alpha _{k}+\gamma_{k})+i\frac{\epsilon l}{2}%
\alpha _{k}-\frac{\Delta r\eta }{2}(\alpha _{k}-\beta _{k})=0, \\
ir\dot{\beta}_{k}& -r(\xi_{k}\beta _{k}-\gamma_{k})-i\frac{\epsilon r}{2}%
\beta _{k}-\frac{\Delta l\eta ^{\ast }}{2}(\beta _{k}-\alpha _{k})=0.
\end{align}
\label{dy}
\end{subequations}
On the basis of the kinetic equations, the time-evolving state of the system and its associated properties can be systematically determined.

In our simulations, we set $M=20{,}000$ and $\xi_{\max }=4\omega _{c}$,
giving $T_{\text{p}}=10{,}000\pi $, thus ensuring that the dynamics are computed well  beyond recurrence effects. The convergence has been verified for these parameters, and the dynamical results are presented in Sec.~\ref{dynamics}.

\section{Results and Discussion}
\label{Result and Discussion}
\subsection{Eigenvalue Spectrum and Exceptional Points}
\label{Energy Spectra}

\begin{figure}[htbp]
\includegraphics[width=8.6cm]{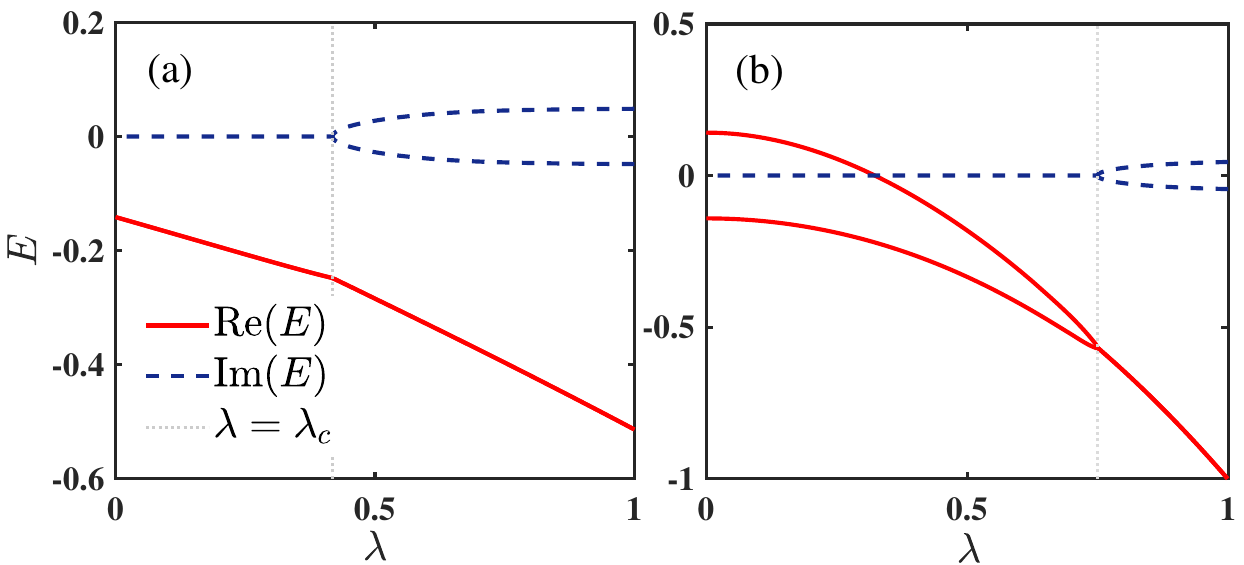}
\vspace{-0.1cm}
\caption{Eigenvalue spectrum of the $\mathcal{PT}$-symmetric non-Hermitian spin-boson model (a) and the corresponding quantum Rabi model (b). Parameters: $\Delta = 0.3$ and $\epsilon = 0.1$.}
\label{Fig1}
\end{figure}

Building on the framework established in Sec.~\ref{Projection Method}, we now analyze the
spectrum of the non-Hermitian spin-boson model. As shown in Fig.~\ref{Fig1}%
(a), increasing the coupling strength drives the system from the $\mathcal{PT}$-unbroken region to the $\mathcal{PT}$-broken region via an EP point. In the $\mathcal{PT}$-broken regions, a pair of complex conjugate eigenvalues emerges, consistent with recent studies on the
single-mode $\mathcal{PT}$-symmetric quantum Rabi model~\cite{Lu2023PRA_108_053712,Wang2025PRA_112_043704}.

Surprisingly, the real eigenvalues in the $\mathcal{PT}$-unbroken region behave quite differently. In contrast to the $\mathcal{PT}$-symmetric quantum Rabi model with a mode $\omega_0=1$, where the two lowest eigenvalues are well separated in Fig.~\ref{Fig1}(b), the present $\mathcal{PT}$-symmetric spin-boson model
exhibits only a single lowest real eigenvalue. This pattern resembles that of its Hermitian counterpart shown in Fig.~\ref{Fig7}(a) of Appendix~\ref{Appen-H}, although its first derivative shows a discontinuity.

Notably, even in the generalized $\mathcal{PT}$-symmetric quantum Rabi model comprising a finite number of bosonic modes, the two lowest real eigenvalues are present. They coalesce at the exceptional point and, upon further increase of the coupling, transition into a pair of complex conjugate  eigenvalues. This phenomenon is analyzed in the Appendix~\ref{Finite Number} (see Fig.~\ref{Fig6}), showing excellent agreement with exact diagonalization results.

This unexpected behavior in the $\mathcal{PT}$-symmetric spin‑boson model stems from the continuum of bosonic modes that extends to zero frequency. Due to this spectral continuity, the energy gap closes in the infrared, allowing successive frequencies $\omega_n$ and $\omega_{n-1}$ to become arbitrarily close. These quasi‑degenerate modes are conveniently indexed by $n$, effectively forming a continuous spectrum above the lowest eigenstate. As a result, information about individual excited states is lost, and only the lowest eigenvalues can be clearly resolved, consistent with the spectrum shown in Fig.~\ref{Fig1}(a).

By exploiting the $\mathcal{PT}$-symmetry of the system, the two imaginary branches emerging in the $\mathcal{PT}$-broken phase can also be identified. This indicates that, for a continuous spectrum~\cite{Naimark1960AMS_16_103,Ashida2020AP_69_249-435}, only a single real eigenvalue branch remains visible. The EP marks the boundary at which the real part of the lowest eigenvalue exhibits a singular cusp, accompanied by a discontinuity in its first derivative. Beyond the EP, pairs of complex‑conjugate eigenvalues appear, mirroring the behavior familiar from non-Hermitian models with finite modes.

\begin{figure}
\includegraphics[width=8.6cm]{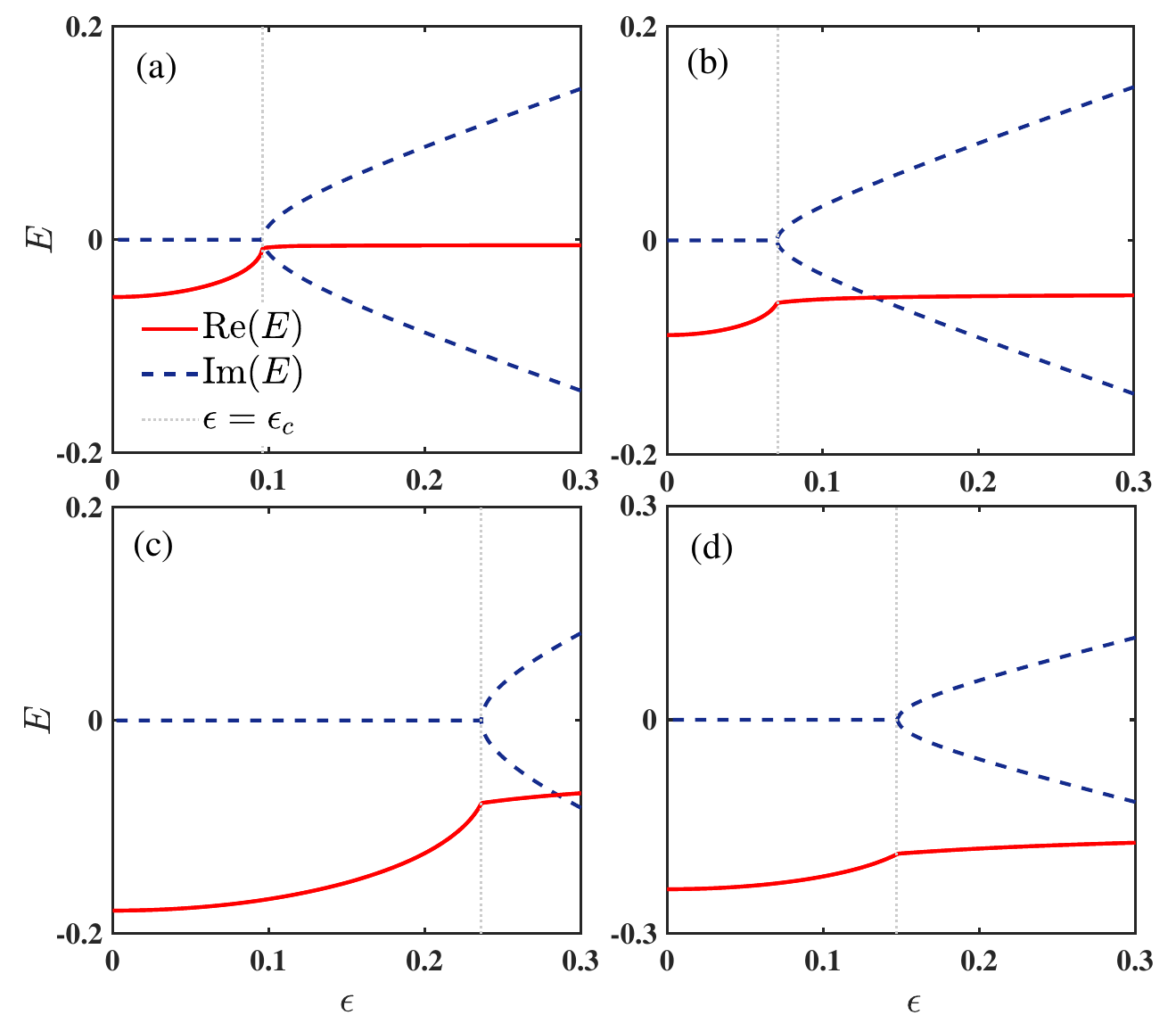}
\vspace{-0.1cm}
\caption{Eigenvalue spectrum of the $\mathcal{PT}$-symmetric non‑Hermitian spin‑boson model versus the bias strength $\epsilon$. Other parameters are set as (a) $\Delta = 0.1,\lambda = 0.01$; (b) $\Delta = 0.1,\lambda = 0.1$; (c) $\Delta = 0.3,\lambda = 0.1$; (d) $\Delta = 0.3,\lambda = 0.3$. }
\label{Fig2}
\end{figure}

    The bias term captures the non-Hermitian character of this system. In the absence of coupling, an EP can be directly obtained at $\epsilon_c = \Delta$ for the non-Hermitian two-level system described by Hamiltonian $H_{\text{TLS}} = -\frac{\Delta}{2}\sigma_x + i\frac{\epsilon}{2}\sigma_z$. This naturally leads to the question: How does the coupling between the two-level system and the bosonic bath alter the exceptional point?

To address this issue, we vary the energy bias $\epsilon$ and systematically examine the spectral properties of the $\mathcal{PT}$-symmetric spin‑boson model. As illustrated in Fig.~\ref{Fig2}, $\mathcal{PT}$-symmetry breaking occurs as the bias $\epsilon$ increases, accompanied by a pronounced change in the spectrum around the critical bias strength $\epsilon_c$, which corresponds to an EP. The real part of the eigenvalue bends in the EP. Beyond this point, a pair of complex-conjugated eigenvalues emerges, as expected.

Specifically, when the coupling strength $\lambda$ is very small (e.g. $\lambda=0.01$), the critical point $\epsilon_c$ for $\mathcal{PT}$-symmetry breaking appears near the bare atomic frequency $\Delta$, as illustrated in Fig.~\ref{Fig2}(a). This suggests that the extremely weak coupling has a negligible effect on the EP of the non-Hermitian two-level system, where $\epsilon_c$ equals $\Delta$. As the coupling enters the ultra-strong coupling regime, for instance, at $\lambda=0.1$ in Fig.~\ref{Fig2}(b), the critical value $\epsilon_c$ shifts to below $\Delta$.  This reduction indicates that $\mathcal{PT}$ symmetry breaking occurs at a lower $\epsilon_c$, implying that the coupling to the bosonic bath effectively renormalizes the frequency of the two-level system. Consequently, the system becomes more susceptible to $\mathcal{PT}$-symmetry breaking and  begins to exhibit characteristic non‑Hermitian behavior (i.e., the emergence of imaginary eigenvalues) at an earlier stage.

As the coupling strength increases further, the critical bias $\epsilon_c$ for $\mathcal{PT}$-symmetry breaking falls markedly, as shown in Figs.~\ref{Fig2}(c)–(d). By contrast, increasing the atomic frequency $\Delta$ requires a larger bias $\epsilon$ to induce $\mathcal{PT}$-symmetry breaking, evident from the comparison between Figs.~\ref{Fig2}(b) and (c). Together, these results systematically illustrate the eigenvalue spectrum features of the non‑Hermitian spin‑boson system, highlighting how both the coupling strength and the energy bias govern its static properties.

\subsection{Dynamics Induced by $\mathcal{PT}$ Symmetry Effects}
 \label{dynamics}
In the previous section, we have showed that the $\mathcal{PT}$-symmetric spin–boson model exhibits markedly different spectral characteristics in the $\mathcal{PT}$-unbroken and $\mathcal{PT}$-broken phases. Based on Eq.~\eqref{dy} obtained from the Dirac–Frenkel TDVP approach, we now proceed to study the dynamical behavior of the non-Hermitian spin–boson system starting from the initial state $\left\vert\Psi (0)\right\rangle=\left\vert0\right\rangle \left\vert \uparrow \right\rangle $, with special attention to the role of the $\mathcal{PT}$ symmetry and to the contrasting dynamical behaviors in the $\mathcal{PT}$-unbroken and $\mathcal{PT}$-broken regimes.

\subsubsection{Spin Dynamics}
\begin{figure}
\includegraphics[width=8.6cm]{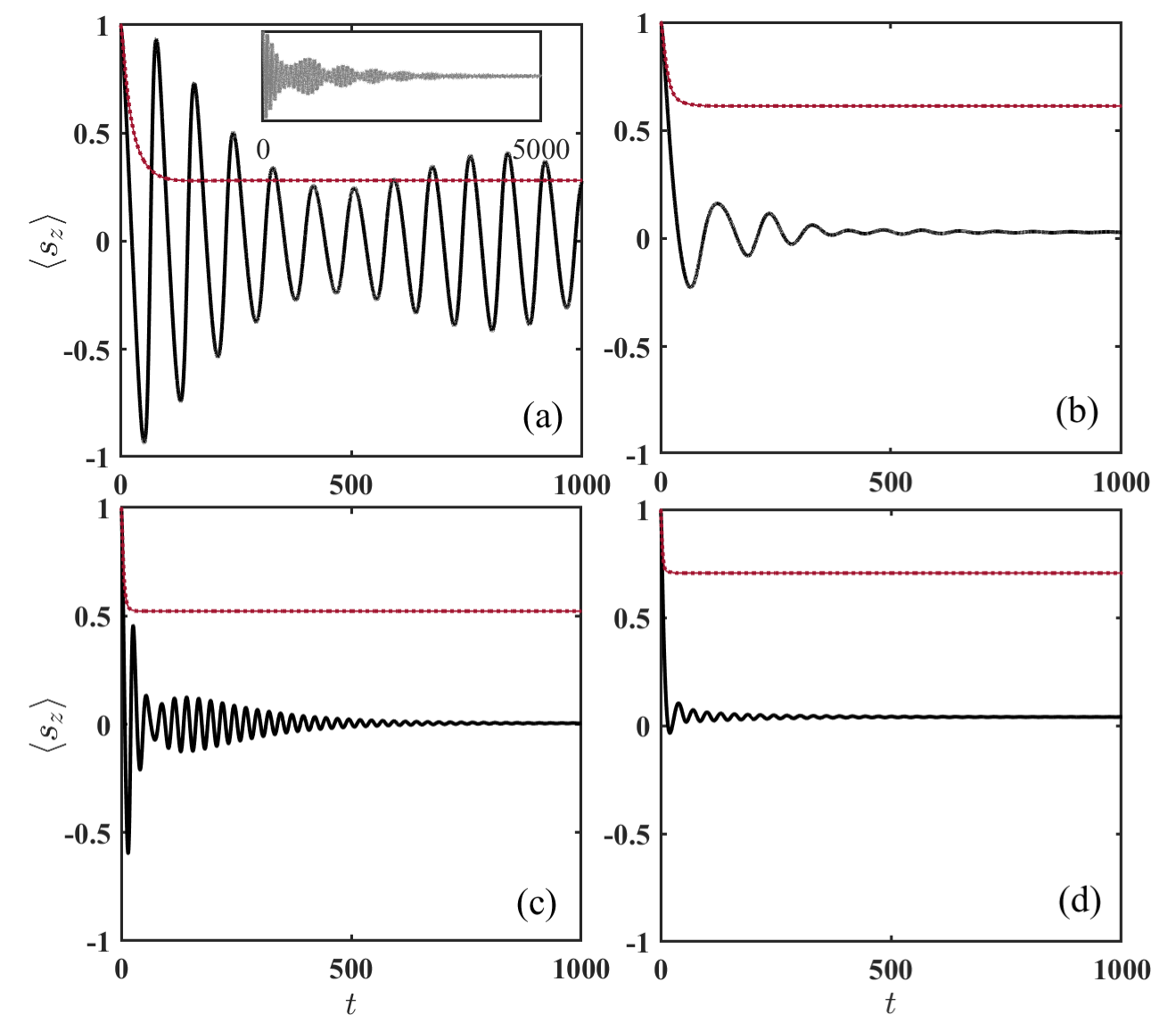}
\vspace{-0.1cm}
\caption{Time evolution of the average spin component $\langle s_{z} \rangle$ under different parameter sets:
(a) $\Delta = 0.1$, $\lambda = 0.01$, $\epsilon = 0.05$ (black), $0.1$ (red);
(b) $\Delta = 0.1$, $\lambda = 0.1$, $\epsilon = 0.05$ (black), $0.1$ (red);
(c) $\Delta = 0.3$, $\lambda = 0.1$, $\epsilon = 0.1$ (black), $0.3$ (red);
(d) $\Delta = 0.3$, $\lambda = 0.3$, $\epsilon = 0.1$ (black), $0.3$ (red).}
\label{Fig3}
\end{figure}

The expectation value of the spin operator $\langle s_z \rangle$, which quantifies the population difference between the spin-up and spin-down states, is defined as
\begin{equation}
\left\langle s_{z}\right\rangle =\frac{\left\vert l\right\vert
^{2}-\left\vert r\right\vert ^{2}}{\left\vert l\right\vert ^{2}+\left\vert
r\right\vert ^{2}}. 
\end{equation}
In Fig.~\ref{Fig3}, we show several dynamical evolutions of $\left\langle s_{z}\right\rangle$ in both $\mathcal{PT}$-unbroken and $\mathcal{PT}$-broken phases. Guided by the spectrum in Fig.~\ref{Fig2}, we select two representative bias values $\epsilon$ for each parameter set, one corresponding to the $\mathcal{PT}$-unbroken regime and the other to the $\mathcal{PT}$-broken regime. In each panel, the black curve corresponds to the dynamics in the $\mathcal{PT}$-unbroken phase, while the red curve represents the $\mathcal{PT}$-broken phase.

In Fig.~\ref{Fig3}(a), where the coupling is weak ($\lambda=0.01$), the spin dynamics in the $\mathcal{PT}$-unbroken regime (black curve) exhibit periodic oscillations with modulated amplitude. Interestingly, this behavior somewhat resembles  the collapse–revival dynamics of the Jaynes–Cummings model~\cite{Scully1997__,Meystre2021__}. Here, although a complete `collapse' does not occur, the oscillations exhibit a transition, with their amplitude first decreasing noticeably and then increasing significantly. The long-term evolution shown in the inset reveals a cycle of amplitude amplification and reduction that persists over an extended period. In contrast, the dynamics in the $\mathcal{PT}$-broken regime (red curve) rapidly approach a steady state with negligible oscillations and almost constant spin occupation, a consequence of strong decoherence induced by the non‑Hermitian bias term.

In Fig.~\ref{Fig3}(b), when the coupling strength reaches the ultrastrong regime ($\lambda = 0.1$), the oscillation frequency and the peak amplitude both decrease monotonically, leading the system to ultimately settle near zero in the $\mathcal{PT}$-unbroken region. Conversely, in the $\mathcal{PT}$-broken phase, strong decoherence rapidly drives the system to a steady state with a higher spin occupation, underscoring the pronounced dynamical difference between the two phases.

Keeping the coupling strength at $\lambda = 0.1$ and the bias at $\epsilon=0.1$, the system, which was initially in the $\mathcal{PT}$-broken phase, can be brought back into the $\mathcal{PT}$-unbroken phase by increasing the tunneling parameter $\Delta$ from $0.1$ to $0.3$, as seen from Fig.~\ref{Fig2}(b) and (c). Dynamically, this spectral transition manifests as a rapid approach to a steady state in the $\mathcal{PT}$-broken phase (Fig.~\ref{Fig3}(b), red curve), contrasting with an oscillatory regime characterized by amplitude amplification in the $\mathcal{PT}$-unbroken phase (Fig.~\ref{Fig3}(c), black curve). When the bias $\epsilon$ is further increased to match $\Delta$ in Fig.~\ref{Fig3}(c), the system re‑enters the $\mathcal{PT}$-broken regime; the dynamics again quickly settles to a steady state with only small residual oscillations, confirming the return to the $\mathcal{PT}$‑broken phase.

Finally, when the coupling strength is increased further to $\lambda = 0.3$ in Fig.~\ref{Fig3}(d), both the oscillation frequency and peak amplitude in the $\mathcal{PT}$-unbroken phase diminish further, and the system stabilizes at an earlier time. Compared with the case in Fig.~\ref{Fig3}(c), the mean spin value $\langle s_z \rangle$ shows a marked rise upon entering the $\mathcal{PT}$-broken regime, underscoring the qualitative contrast in dynamics between different coupling strength. Together, these results illustrate how the interplay among coupling strength, tunneling frequency, and energy bias drives transitions between the two phases, influences decoherence, and gives rise to the distinctive dynamical signatures observed in the system.

To clarify the distinctions between the $\mathcal{PT}$-symmetric spin-boson model and its Hermitian counterpart, we also investigate the spin dynamics of the Hermitian spin–boson model, as shown in Fig.~\ref{Fig7}(b) of Appendix~\ref{Appen-H}. At weak coupling ($\lambda = 0.01$), the system no longer exhibits any oscillations with amplitude amplification over time. As the bias $\epsilon$ increases from $0.05$ to $0.1$, the dynamics becomes more oscillatory—in sharp contrast to the rapid stabilization observed in the non‑Hermitian case in Fig.~\ref{Fig3}(a).

This behavior underscores a key difference from the non‑Hermitian dynamics, where $\mathcal{PT}$ symmetry, before its breaking, plays a crucial role in restricting the flow of information into the environment, thus suppressing decoherence and extending relaxation times~\cite{Duarte2018E_121_50006,Dey2019PLA_383_125931,Naghiloo2019NP_15_1232-1236,Duttatreya2026AoP_484_170298a}. Once the $\mathcal{PT}$ symmetry is broken, a dissipative channel effectively opens, allowing information to leak into the environment more rapidly and driving the system quickly to a steady state~\cite{Naghiloo2019NP_15_1232-1236}. This phenomenon illuminates an intrinsic feature of non‑Hermitian dynamical systems.

\subsubsection{Dynamics of Photon Occupation and Normalization Factor}

\begin{figure}[tbp]
\includegraphics[width=8.6cm]{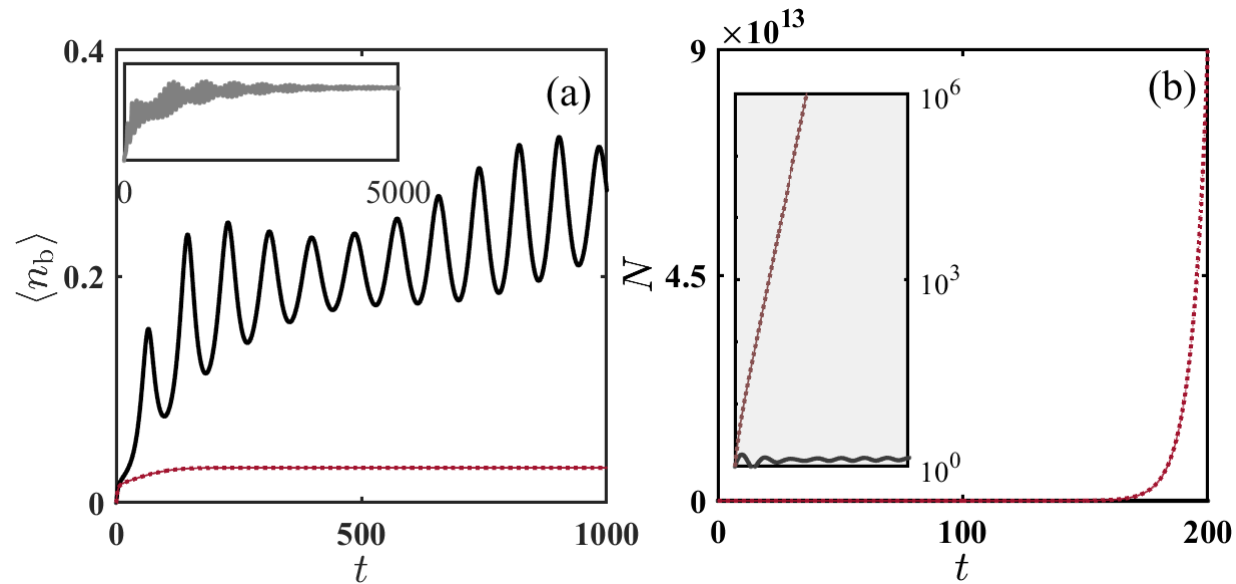} \vspace{-0.1cm}
\caption{ (a) Time evolution of the average photon occupation $\langle n_{\text{b}}\rangle $ and (b) dynamics of the normalization factor $\mathcal{N}$. The inset in (b) displays the same data on a exponential $y$-axis. Parameters: (a) $\Delta = 0.1$, $\protect\lambda =
0.01 $, $\protect\epsilon = 0.05$ (black), $0.1$ (red); (b) $\Delta = 0.3$, $%
\protect\lambda = 0.1$, $\protect\epsilon = 0.1$ (black), $0.3$ (red). }
\label{Fig4}
\end{figure}

The average photon number $\langle n_{\text{b}}\rangle =\sum_{k}\langle
a_{k}^{\dagger }a_{k}\rangle $ in this system is given by
\begin{equation}
\langle n_{\text{b}}\rangle =\frac{|l|^{2}\sum_{k}|\alpha
_{k}|^{2}+|r|^{2}\sum_{k}|\beta _{k}|^{2}}{|l|^{2}+|r|^{2}}.
\end{equation}%
In Fig.~\ref{Fig4}, the black curves correspond to  parameters in the $%
\mathcal{PT}$-unbroken phase, and the red curves to the $\mathcal{PT}$%
-broken phase. The dynamics of photon occupancy before and after $\mathcal{PT%
}$ symmetry breaking shows a clear contrast, as illustrated in Fig.~\ref%
{Fig4}(a) for $\lambda =0.01$. In the $\mathcal{PT}$-unbroken  region, the
photon number exhibits oscillatory behavior. To highlight this, an inset is
included in the figure, displaying the long-time evolution of the
photon occupancy. In particular, the system exhibits pronounced oscillations with periodic amplified amplitude. By contrast, in the $\mathcal{PT}$-broken phase, enhanced decoherence rapidly damps the oscillations and drives the system to a steady state. This demonstrates that symmetry breaking significantly modifies the dissipative properties of the system, notably accelerating its relaxation toward equilibrium, in agreement with the behavior of $\langle s_{z}\rangle$ reported in Ref.~\cite{Naghiloo2019NP_15_1232-1236}.

The normalization factor, which ensures proper normalization of the
system's eigenstate, is defined as
\begin{equation}
\mathcal{N}=\langle \Psi (t)|\Psi (t)\rangle =|l|^{2}+|r|^{2}.
\end{equation}%
Since $\frac{d\mathcal{N}}{dt}=(|l|^{2}-|r|^{2})\epsilon \neq 0$, the normalization
factor is not conserved, which is a direct signature of the non-Hermitian
character of the system, unlike Hermitian systems where $\frac{d\mathcal{N}}{dt}=0$. This behavior is illustrated in the inset of Fig.~\ref{Fig4}(b), where $\mathcal{N}$ increases over time in both the $\mathcal{PT}$-unbroken and $\mathcal{PT}$%
-broken phases. Notably, $\mathcal{N}$ increases much more rapidly in the $\mathcal{PT}
$-broken phase, indicating that strong decoherence drives the system away
from its initial state and accelerates its approach to a steady state~\cite{Naghiloo2019NP_15_1232-1236}. In certain parameter regimes of the $\mathcal{PT}$-unbroken phase, however, $\mathcal{N}$ grows only gradually, suggesting that the
$\mathcal{PT}$-unbroken phase can maintain  a dynamically balanced state with suppressed decoherence despite the non-Hermitian nature~\cite{Duarte2018E_121_50006,Dey2019PLA_383_125931,Naghiloo2019NP_15_1232-1236,Duttatreya2026AoP_484_170298a}.

Based on the preceding analysis, it is clear that in the $\mathcal{PT}$-symmetric non-Hermitian spin-boson model,  the dynamics of the spin, boson occupation, and normalization factor differ substantially between the $\mathcal{PT}$-unbroken and $\mathcal{PT}$-broken phases. These observations highlight  several key points: In the $\mathcal{PT}$-unbroken phase, the system approaches its steady state more slowly than its Hermitian counterpart, reflecting the decoherence-suppressing role of $\mathcal{PT}$ symmetry. Conversely, in the $\mathcal{PT}$-broken phase, additional dissipation channels open, inducing rapid decoherence. These dynamical features are consistent with the spectral analysis in Sec.~\ref{Energy Spectra}, where the spectrum remains real and is protected by $\mathcal{PT}$ symmetry before it breaks, but becomes complex, accompanied by enhanced dissipation, once the symmetry is broken.

\section{Conclusions and Outlook}
\label{Conclusion}
In this paper, we explore several key properties of the $\mathcal{PT}$-symmetric non-Hermitian spin-boson model, with a focus on its eigenvalue spectrum and dynamical behavior.

Regarding the eigenvalue spectrum, the $\mathcal{PT}$-symmetric
non-Hermitian spin-boson model undergoes a transition from the $\mathcal{PT}$-unbroken phase to the $\mathcal{PT}$-broken phase through an EP. In the $\mathcal{PT}$-broken region, a pair of complex-conjugated eigenvalues emerges, while the $\mathcal{PT}$-unbroken region is characterized by a single lowest real eigenvalue. This behavior stands in sharp contrast with that of its single-mode counterpart, i.e., the $\mathcal{PT}$-symmetric quantum Rabi model, which exhibits two well-separated lowest eigenvalues. Moreover, in the latter, there exist infinitely many EPs, each involving the coalescence of two discrete eigenvalues. However, because of the continuous nature of its energy spectrum, the non-Hermitian spin-boson model resembles its Hermitian counterpart in possessing only one well-defined ground state.

Turning to the dynamical properties, we examine the spin population, photon occupation, and normalization factor. In the $\mathcal{PT}$-unbroken phase, the system exhibits an oscillatory behavior with slower relaxation, reflecting suppressed decoherence and a prolonged approach to the steady state. In contrast, in the $\mathcal{PT}$-broken phase, additional dissipation channels accelerate the decoherence, allowing the system to stabilize rapidly. At small coupling strengths, the oscillations with periodic amplitude amplification emerge in the $\mathcal{PT}$-symmetric model, a feature absent in its Hermitian counterpart. The normalization factor oscillates in the $\mathcal{PT}$-unbroken phase but grows rapidly in the $\mathcal{PT}$-broken phase, further illustrating the enhanced decoherence in the latter region.

In summary, our study demonstrates the crucial role of $\mathcal{PT}$-symmetry in governing both the eigenvalue spectrum and the dynamical properties of non-Hermitian spin-boson models. Using ansatze adapted from Hermitian frameworks, we successfully captured qualitative distinctions between different phases. While both static and dynamical approaches can be further refined---for instance, through the use of multi-coherent states---to achieve quantitative accuracy, the fundamental qualitative features of the model remain consistent. This work advances the understanding of eigenvalue spectra and dynamical behavior in such many-body $\mathcal{PT}$-symmetric non-Hermitian systems, providing valuable insights for future theoretical and experimental investigations in the field. In future work, we will extend the analysis to other spectral densities to compare how different spectral structures affect the manifestation of $\mathcal{PT}$ symmetry. We also plan to quantitatively examine the differences between the effective non-Hermitian treatment adopted here and Lindblad-type open-system approaches, with particular emphasis on qualitative distinctions and the balance between accuracy and computational cost.
\begin{acknowledgments}
This work is supported by the National Key R\&D Program of China under Grant No. 2024YFA1408900 and the National Natural Science Foundation of China under Grant No. 92565201.
\end{acknowledgments}
\appendix

\section{Finite Multi-Mode Non-Hermitian Spin-Boson Model}
\label{Finite Number}
\begin{figure}[h]
\includegraphics[width=8.6cm]{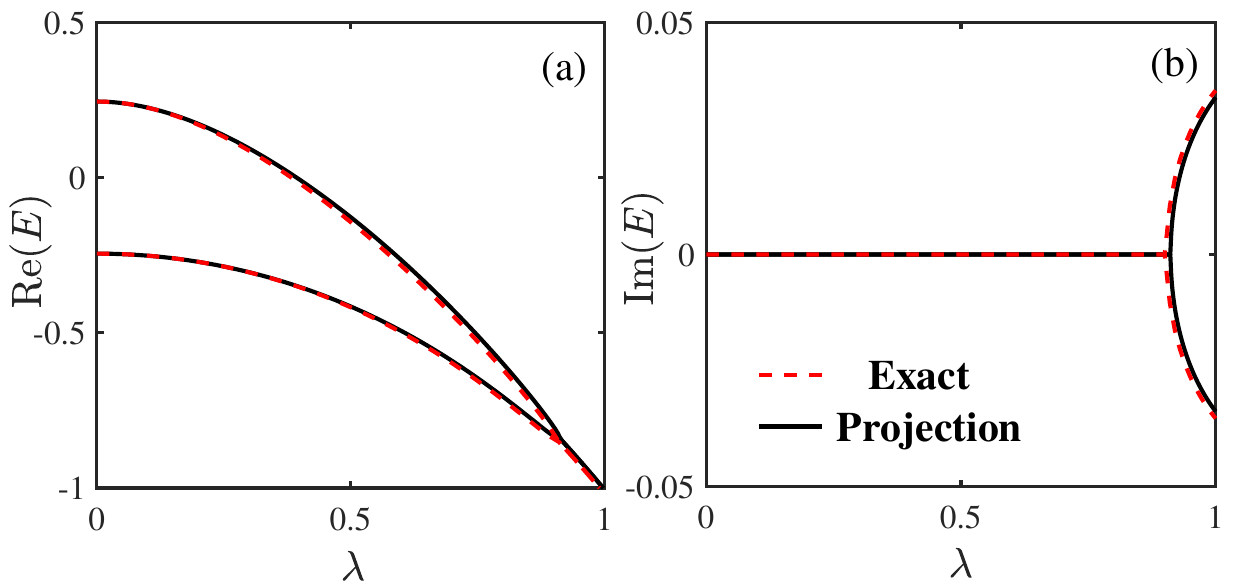} \vspace{-0.1cm}
\caption{Lowest two eigenvalues of the non‑Hermitian quantum Rabi model. (a) Real parts; (b) Imaginary parts. Parameters: $\omega_0=1$, $\Delta = 0.5$, $\epsilon = 0.1$.}
\label{Fig5}
\end{figure}

The non-Hermitian quantum Rabi model describes the simplest form of a multi-mode $\mathcal{PT}$-symmetric spin-boson model, constrained to a single bosonic mode~\cite{Lu2023PRA_108_053712, Wang2025PRA_112_043704}. The Hamiltonian of the system is given by
\begin{equation}
H_{\text{Rabi}} = -\frac{\Delta}{2} \sigma_x + i\frac{\epsilon}{2} \sigma_z + \omega_0 a^{\dagger} a + \lambda \sigma_z (a + a^{\dagger}).
\end{equation}

We again employ the projection method established in Sec.~\ref{Projection Method} to compute the lower eigenvalues for comparison. As shown in Fig.~\ref{Fig5}, the two lowest eigenvalues obtained from the projection method agree well with those from numerical exact diagonalization. Both methods consistently identify the EPs where the two lowest real eigenvalues coalesce at a critical coupling strength $\lambda_c$. This outcome provides robust support for studying more general non-Hermitian spin‑boson systems.  Higher eigenvalues are not discussed here.

The projection method can also be applied to the $\mathcal{PT}$-symmetric spin-boson model with $M$ discrete bosonic modes. The corresponding Hamiltonian is given by
\begin{equation}
H_{M}=-\frac{\Delta }{2}\sigma _{x}+i\frac{\epsilon }{2}\sigma
_{z}+\sum_{n=1}^{M}\omega_{n}a_{n}^{\dag }a_{n}+\sigma
_{z}\sum_{n=1}^{M}g_{n}(a_{n}+a_{n}^{\dag }).  \label{M}
\end{equation}
Without loss of generality, the mode frequencies are  taken to be equally spaced, expressed as $\omega_n = \omega_1 + (n-1)\delta$, $\delta = (\omega_M - \omega_1)/(M-1)$ for $n = 1,2,\dots,M (M>1)$, where $\omega_1$ and $\omega_M$ represent the lowest and highest frequencies, respectively. Throughout the following analysis, we fix $\omega_1 = 1$ and $\omega_M = 1.4$. The corresponding coupling strengths are defined as $g_n = \sqrt{\lambda\omega_n/(M-1)}$. 

\begin{figure}[h]
\includegraphics[width=8.6cm]{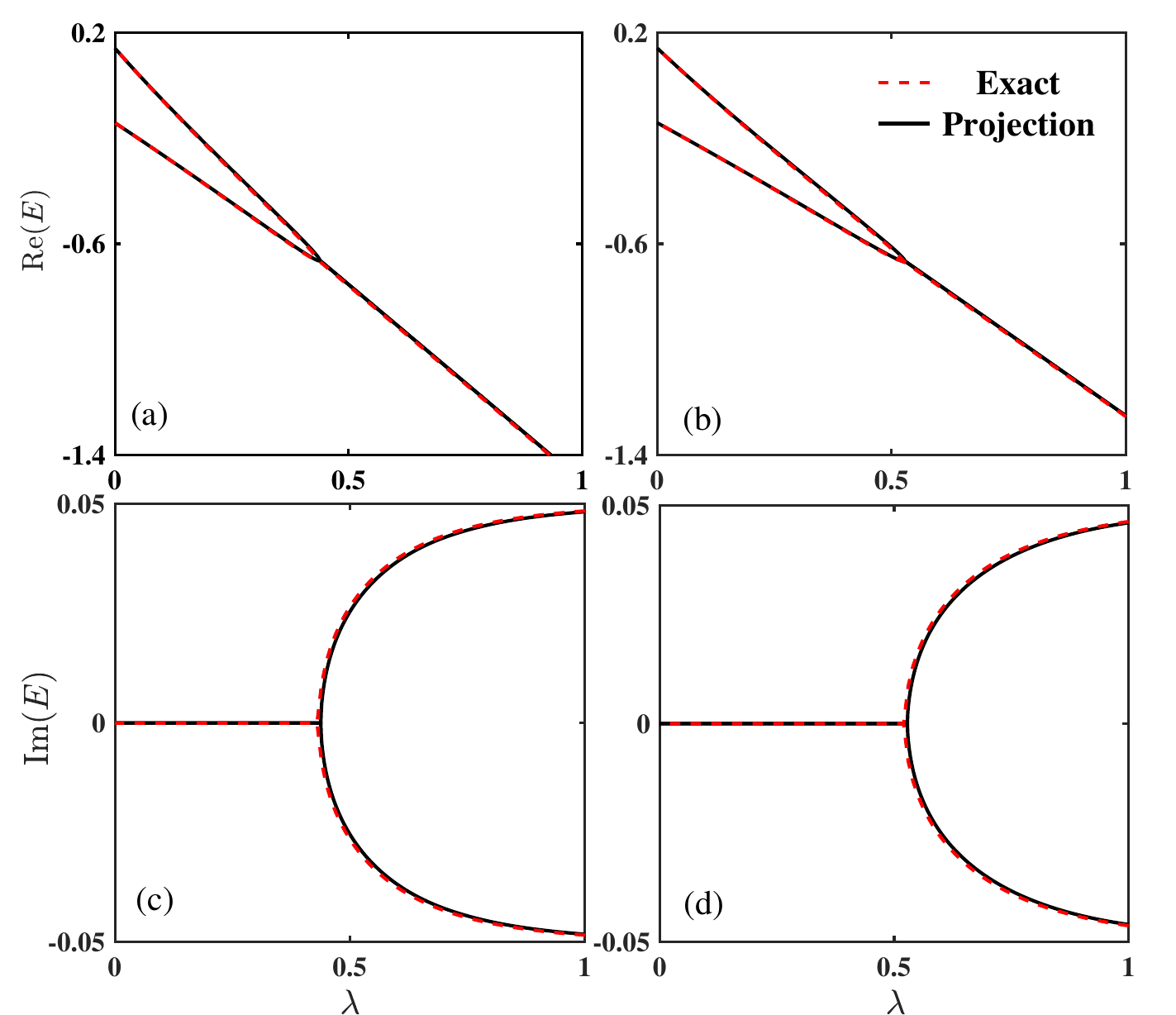} \vspace{-0.1cm}
\caption{Lowest two eigenvalues of the $\mathcal{PT}$-symmetric non-Hermitian spin-boson model with a finite number of bosonic modes. Parameters: $\Delta = 0.3$, $\epsilon = 0.1$, for $M=3$ (left panel) and $M=5$ (right panel).}
\label{Fig6}
\end{figure}

In Fig.~\ref{Fig6}, we plot the lowest two eigenvalues of the system for different numbers of bosonic modes $M$ (higher eigenvalues are omitted). The results from the projection method and exact diagonalization show excellent agreement, further validating the reliability and effectiveness of the ansatz employed in the projection method for capturing the properties of the non-Hermitian system. In this finite multi-mode non-Hermitian spin-boson model, we also observe spectral singularities: as the coupling strength increases, the two real eigenvalues coalesce at the EP, while previously absent imaginary parts of the eigenvalue emerge beyond this point. This behavior is consistent with that of the non-Hermitian quantum Rabi model, suggesting that the finite multi-mode spin-boson model undergoes a phase transition analogous to $\mathcal{PT}$-symmetry breaking in the single-mode case. 
\section{Energy Spectra and Dynamical Properties of the Hermitian Spin--Boson Model}
\label{Appen-H}

The Hermitian spin-boson model consists of a two-level qubit coupled to a continuous bosonic bath, which is given by
\begin{equation}
H_{\text{H}}=-\frac{\Delta }{2}\sigma _{x}+\frac{\epsilon }{2}\sigma
_{z}+\sum_{j}\omega _{j}a_{j}^{\dag }a_{j}+\sigma
_{z}\sum_{j}g_{j}(a_{j}+a_{j}^{\dag }). 
\label{HH}
\end{equation}

The parameters are defined in the same way as those in its non-Hermitian counterpart in Eq.~\eqref{H}, except for the real bias term $\epsilon$. However, the relation $[\mathcal{PT},H_{\text{H}}]\neq0$, indicating the loss of the $\mathcal{PT}$-symmetry. In this Appendix, we present the ground-state energy and dynamical properties of the Hermitian spin-boson model and compare them with those of the $\mathcal{PT}$-symmetric non-Hermitian model in Eq.~\eqref{H} for the same bias strength $\epsilon$.

\begin{figure}[thbp]
\includegraphics[width=8.5cm]{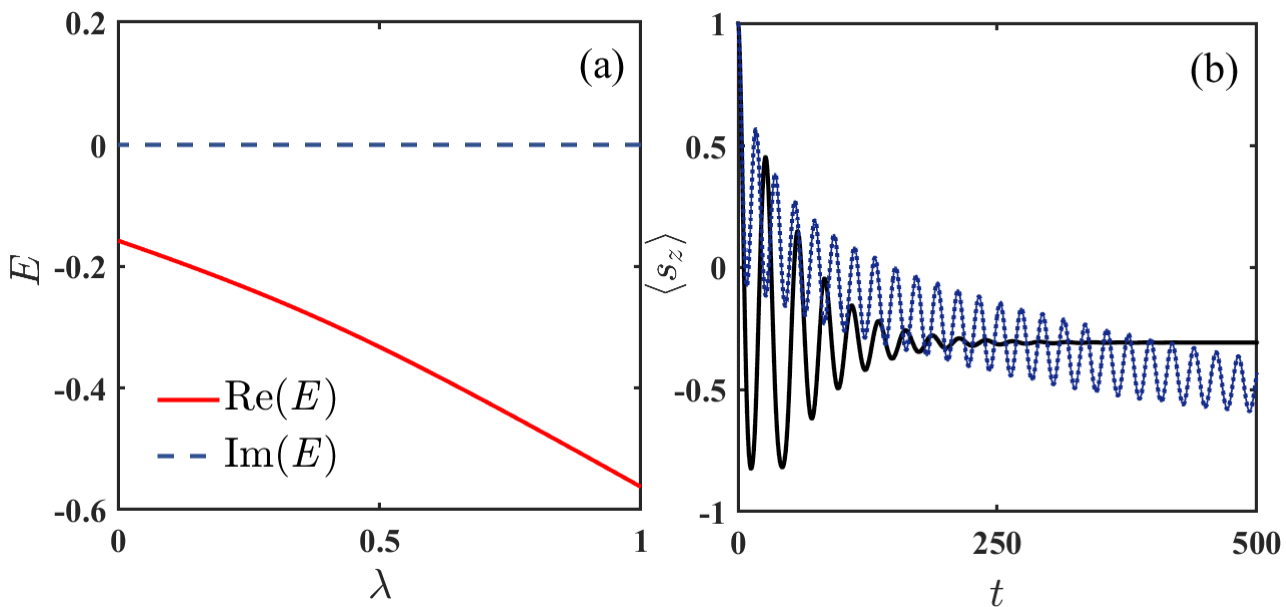}
\vspace{-0.1cm}
\caption{Hermitian spin-boson model: (a) Spectrum versus coupling strength $\lambda$. (b) Time evolution of the average spin component $\langle \sigma_z \rangle$. Parameters: (a) $\Delta = 0.3, \epsilon = 0.1$; (b) $\Delta =0.1, \epsilon =$ 0.05 (black), 0.1 (blue), $\lambda = 0.01$.}
\label{Fig7}
\end{figure}
Following the methods outlined in Sec.~\ref{Model and Method} for the non-Hermitian framework, we also present the energy spectrum of the Hermitian spin-boson model, as shown in Fig.~\ref{Fig7}(a). We observe that the ground-state energy  varies continuously with the coupling strength, in sharp contrast to the non-Hermitian case under identical parameters, displayed in Fig.~\ref{Fig1}(a). In Fig.~\ref{Fig7}(b), we present the spin dynamics of the Hermitian spin-boson model using the same parameter settings as in Fig.~\ref{Fig3}(a). Under a small bias of $\epsilon = 0.05$ and weak coupling $\lambda = 0.01$, the system oscillates briefly before rapidly settling into a steady state with negative spin occupation, without exhibiting amplitude amplification. In contrast, for a larger bias of $\epsilon =0.1$, the system exhibits oscillations with a higher frequency but a smaller amplitude, which persist over a long duration without converging to a steady state. This is completely different from the dynamics of the corresponding non-Hermitian system.

%

\end{document}